\documentclass[11pt]{article}
\usepackage[utf8]{inputenc}
\usepackage{amsmath}
\usepackage{graphicx}
\usepackage{url}
\usepackage{setspace}

\title{Religious affiliation, education and Internet use}
\author{Allen B. Downey\footnote{Professor of Computer Science, Olin College of
    Engineering, Needham MA 02492.  Email: {\tt
      allen.downey@olin.edu}, Web: {\tt allendowney.com}}}
\date{March 2014}

\begin{document}

\maketitle 

\begin{abstract}

Using data from the General Social Survey, we measure the
effect of education and Internet use on religious affiliation.

We find that Internet use is associated with decreased probability
of religious affiliation; for moderate use (2 or more hours per week)
the odds ratio is 0.82 (CI 0.69--0.98, $p=0.01$).  For heavier use
(7 or more hours per week) the odds ratio is 0.58 (CI 0.41--0.81,
$p<0.001$).

In the 2010 U.S. population, Internet use could account for 5.1
million people with no religious affiliation, or 20\% of the observed
decrease in affiliation relative to the 1980s.  Increases in college
graduation between the 1980s and 2000s could account for an
additional 5\% of the decrease.

\end{abstract}

\begin{figure}
\centerline{\includegraphics[width=4in]{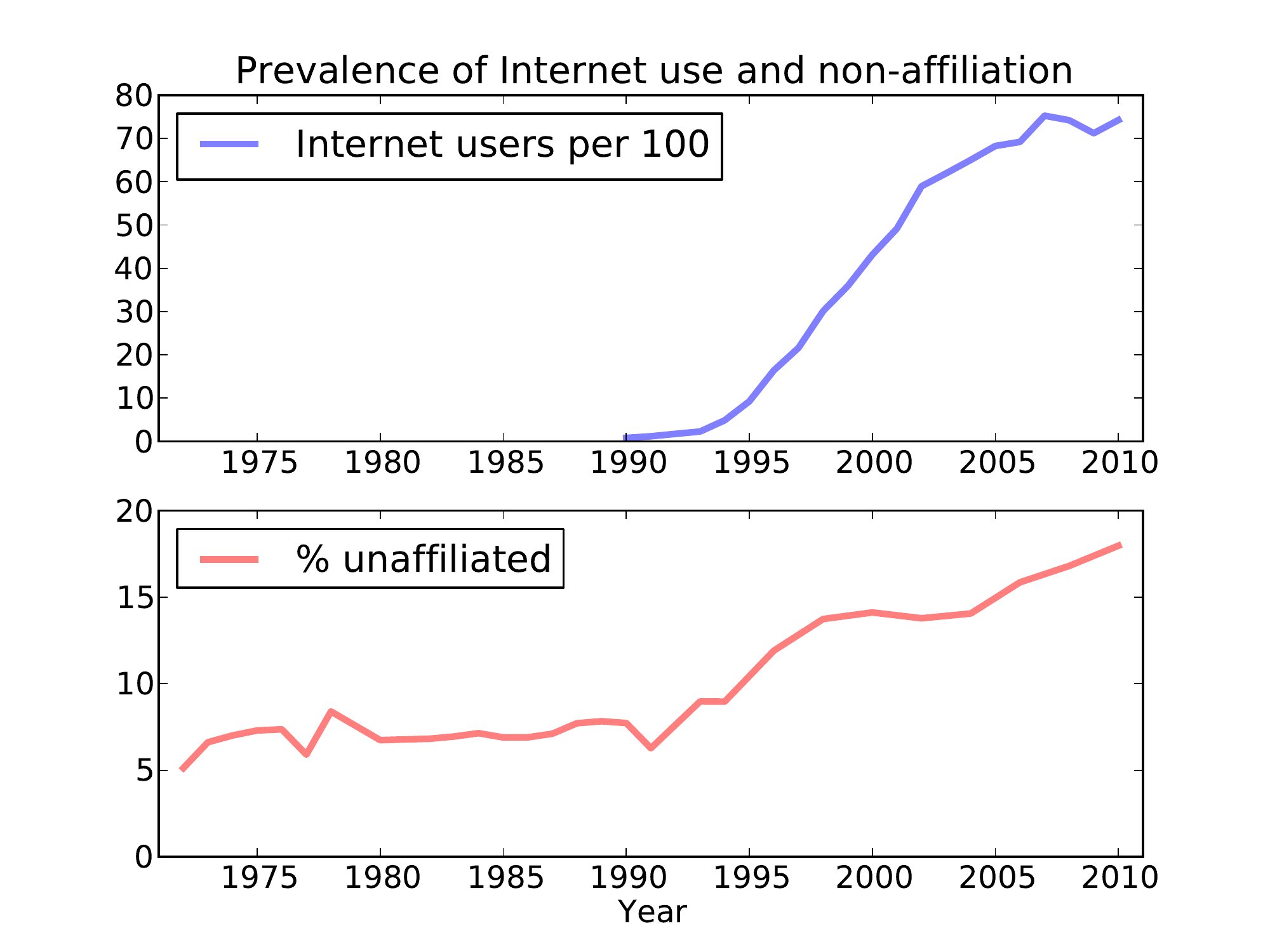}}
\caption{Internet users per 100 people and the fraction of the
  population with no religious preference, 1972--2010.  Sources:
World Bank, General Social Survey (GSS).}
\label{fig.internet}
\end{figure}

From 1990 to 2010 the fraction of people in the United States with no
religious preference increased from 8\% to 18\%, based on data from
the General Social Survey (GSS).  At the same time, the fraction of
Protestants dropped from 62\% to 51\%; the fraction of Catholics and
Jews did not change significantly; the fraction of
other religions increased from 3.3\% to 4.2\%.

During the same period, the prevalence of Internet use increased from
essentially zero to nearly 80\%.  Figure~\ref{fig.internet} shows
Internet users per 100 people and the fraction of the population with
no religious preference.  Also during this period, the fraction
of the population graduating from college increased from 17\% to
27\%.

In this paper we explore the effect of increases in college
education and Internet use on religious affiliation.

Of course there are other factors that might contribute to
religious disaffiliation.  Hout and Fischer investigate this trend
(through 2000) and identify as causes political beliefs and
generational effects, including a decrease in the fraction of people
raised with no religion \cite{hout2002more}.

Smith and Kim study the decline in the Protestant majority,
identifying as causes a decline in ``the intergenerational retention
rate for Protestants, ... shifts in immigration, [and] an increasing
share of people identifying as generic Christians.''
\cite{smith2005vanishing}.

Schwadel studies generational effects, especially people raised with
no religion \cite{schwadel2010period}, and the effect of education
\cite{schwadel2011effects}.

Vargas investigates the effect of ``political attitudes, religious
skepticism, life stressors, and sociodemographic characteristics,''
and finds that ``each is associated with ... leaving religion, but
not necessarily in ... expected directions.''
\cite{vargas2011retrospective}.

Wilcox et al investigate the differential decline among the
high school graduates, relative to college graduates, and identify
as causes ``economic characteristics, current and past family
characteristics, and attitudes toward premarital sex
\cite{wilcox2012no}.

Although prior studies have identified factors associated with
disaffiliation, none have quantified what fraction of the observed
change can be attributed to each factor.  We believe that this
is the first study to show that Internet use is associated with
disaffiliation, and to estimate the magnitude of the effect.

\section{Methodology}

We selected the following variables from the General Social Survey
(GSS), available from \url{http://www3.norc.org/gss+website/}:

\begin{description}

\item[RELIG]: ``What is your religious preference?''

\item[RELIG16]: ``In what religion were you raised?''

\item[AGE]: Respondent's age when surveyed.

\item[YEAR]: Year of survey.

\item[EDUC]: ``What is the highest grade in elementary school or high 
school that you finished and got credit for?''

\item[INCOME]: ``In which of these groups did your total family income, from 
all sources, fall last year before taxes, that is?''

\item[SEI]: Respondent socioeconomic index (computed by GSS based on
  respondent's occupation).

\item[SRCBELT]: Classification of respondent's metropolitan
  statistical area as urban, suburban or rural (coded by GSS based on
  location of interview and U.S. Census data).

\item[WWWHR]: ``Not counting e-mail, about how many minutes or hours per 
week do you use the Web?''

\item[COMPWT]: Respondent's computed sample weight.  All statistics
reported in this paper reflect these weights.

\end{description}

We use data from GSS survey years 2000, 2002, 2004, 2006 and 2010.  In
2008 questions about Internet use were not asked.

These survey years include 14 948 respondents.  By the design of the
GSS, not all respondents are asked all questions.  We excluded
respondents who were not asked or did not answer one or more of the
relevant questions, yielding 8951 respondents.

Using Python programs available from
\url{https://github.com/AllenDowney/internet-religion}, we compute the
following recoded variables:

\begin{description}

\item[has\_relig]: 1 if the respondent reported any religious
  affiliation when interviewed as an adult, or 0 if the respondent
  reported "None".

\item[had\_relig]: 1 if the respondent reported being raised in a
  religion, 0 otherwise.

\item[top75\_income]: 1 if the respondent reports annual household
  income of \$25,000 or more, which is the highest bracket in the survey.
  About 75\% of respondents exceed this threshold.

\item[born\_from\_1960]: year the respondent was born minus 1960
  (subtracting 1960 makes it easier to interpret the results of the
  regression).

\item[educ\_from\_12]: number of years of school completed, minus 12;
  so 0 indicates a high school graduate; 4 or more indicates a college
  graduate.

\item[www2]: 1 if the respondent reports using the Internet 2 of more
  hours per week, 0 otherwise.  53\% of respondents exceed this threshold.

\item[www7]: 1 if the respondent uses the Internet more than 7 hours
  per week, 0 otherwise.  25\% of respondents exceed this threshold.

\end{description}

We compute logistic regressions with \verb"has_relig" as the dependent
variable; \verb"had_relig", \verb"top75_income",
\verb"born_from_1960" and \verb"educ_from_12" are control variables.

We also ran regressions with control variables based on \verb"SEI" and
and \verb"SRCBELT", but they were not statistically significant and
they had negligible effect on the coefficients of the other
variables.

\section{Results}

Table 1 shows results from Model 1, which estimates the effect of
Internet use after controlling for religious upbringing, income, year
born, and education.  For \verb"born_from_1960" the odds ratio is for
someone born in 1970, relative to someone born in 1960.  For
\verb"educ_from_12" the odds ratio is for a college graduate relative
to a high school graduate.

\begin{table}
\begin{tabular}{|l|r|r|r|}\hline
Variable&Odds ratio&Probability&p-value\\
\hline
\verb"(Intercept)"&0.79 (0.65, 0.97)&44 (39, 49)&0.012\\
\verb"had_relig"&11 (8.8, 13)&89 (88, 91)&*\\
\verb"top80_income"&1.3 (1.1, 1.4)&91 (90, 92)&0.002\\
\verb"born_from_1960"&0.8 (0.77, 0.83)&89 (88, 91)&*\\
\verb"educ_from_12"&0.82 (0.74, 0.91)&87 (86, 89)&*\\
\verb"www2"&0.77 (0.66, 0.88)&84 (83, 85)&*\\
\hline
\end{tabular}

\caption{Model 1: odds ratios and cumulative probabilities, with 95\%
  confidence intervals; * indicates statistical significance at
  $p<0.001$}
\end{table}

Odds ratios can be difficult to interpret; the cumulative
probabilities are intended to help.  If we start with a hypothetical
person raised with no religion, with income in the lowest quintile,
born in 1960, with high school education but no college, and low
Internet use (less than 2 hours per week), the probability that this
person has a religious affiliation as an adult is 44\%.  Now we change
one variable at a time:

\begin{itemize}

\item If this person had been raised in some religion, that would
  increase the chance of religious affiliation to 89\%

\item If this person's income were in the top 75\%, that would
  increase the chance of religious affiliation to 91\%

\item If (in addition) this person were born 10 years later (in 1970)
  the chance would drop to 89\%.

\item If (in addition) this person went to college, the chance would
  drop to 87\%.

\item If (in addition) this person used the Internet 2 or more hours
per week, the chance would drop to 84\%.

\end{itemize}

\subsection{Multiple levels of Internet use}

Internet use and affiliation show a ``dose-response'' relationship
where increasing levels of use yield higher levels of response.

\begin{table}
\begin{tabular}{|l|r|r|r|}\hline
Variable&Odds ratio&Probability&p-value\\
\hline
\verb"(Intercept)"&0.79 (0.65, 0.97)&44 (39, 49)&0.011\\
\verb"had_relig"&11 (8.8, 13)&89 (88, 91)&*\\
\verb"top80_income"&1.3 (1.1, 1.4)&91 (90, 92)&0.002\\
\verb"born_from_1960"&0.8 (0.77, 0.84)&89 (88, 91)&*\\
\verb"educ_from_12"&0.83 (0.74, 0.92)&87 (86, 89)&*\\
\verb"www2"&0.88 (0.75, 1)&86 (85, 87)&0.062\\
\verb"www7"&0.74 (0.64, 0.88)&82 (80, 84)&*\\
\hline
\end{tabular}

\caption{Model 2: odds ratios and cumulative probabilities, with 95\%
  confidence intervals; * indicates statistical significance at
  $p<0.001$}
\end{table}

Model 2, presented in Table 2, includes two levels of Internet use,
\verb"web2" and \verb"web7".  In this model, \verb"web2"
is not statistically significant, but the model as a whole is
significantly better than Model 1 augmented with an additional random
variable (p=0.05).  To compare models, we use the self-information of
partitioning (see Methodological Notes, below).

\subsection{Subgroups}

The effect of Internet use is easiest to demonstrate for people with a
religious upbringing.  

We divide the sample into people raised with and without religion.
Among 686 people raised with no religion, the effect of Internet use
is small and not statistically significant (odds ratio 0.94,
$p=0.76$).  Thus the effect may be smaller, or just
less apparent due to small sample size.

\begin{table}
\begin{tabular}{|l|r|r|r|}\hline
Variable&Odds ratio&Probability&p-value\\
\hline
\verb"(Intercept)"&8.5 (7.3, 10)&90 (88, 91)&*\\
\verb"top80_income"&1.3 (1.1, 1.6)&92 (91, 93)&*\\
\verb"born_from_1960"&0.83 (0.79, 0.86)&90 (89, 92)&*\\
\verb"educ_from_12"&0.8 (0.71, 0.89)&88 (87, 90)&*\\
\verb"www2"&0.82 (0.69, 0.98)&86 (85, 88)&0.01\\
\verb"www7"&0.71 (0.6, 0.83)&81 (80, 83)&*\\
\hline
\end{tabular}

\caption{Model 2 restricted to people raised with some religion: odds
  ratios and cumulative probabilities, with 95\% confidence intervals;
  * indicates statistical significance at $p<0.001$}
\end{table}

Table 3 shows the result of Model 2 applied to 8265 people raised with
some religion.  Both \verb"web2" and \verb"web7" are statistically
significant, and their effect is slightly stronger than in the
combined population.

Combining \verb"web2" and \verb"web7", heavy Internet use has an odds
ratio of 0.58 and, for the hypothetical case above, decreases the
probability of religious affiliation by 7 percentage points.

\section{Explaining changes in disaffiliation}

The prevalence of people with no religious affiliation has increased
substantially since 1990.  Based on GSS respondents partitioned by
decade, the fraction of people in the U.S. with no religious
affiliation was 7.1\% in the 1980s, 10.2\% in the 1990s, and 15.3\% in
the 2000s.  The difference between the 1980s and the 2000s is 8.2
percentage points, or 25 million people in the 2010 population
(309 million according to the U.S. Census).

We can use models from the previous section to estimate how much
of this change can be attributed to each contributing factor:

\begin{itemize}

\item Not surprisingly, the factor with the strongest effect on
  religious affiliation is religious upbringing, and the number
  of people raised without religion is increasing, from
  3.3\% in the 1980s to 5.0\% in the 1990s and 7.7\% in the 2000s.

\item College education decreases the chance of religious affiliation,
  and the prevalence of college eduction is increasing.
  The fraction of people in the U.S. with 16 or more years of
  education was 17.4\% in the 1980s, 24.4\% in the 1990s and 27.2\% in
  the 2000s.

\item Internet use decreases the chance of religious affiliation;
  in the 2010s, 53\% of the population used the Internet at least
  2 hours per week, and 25\% more than 7 hours.  Internet use
  in the 1980s was essentially zero.

\item Even controlling for education and Internet use, there is
  a strong generational effect; people born later are less likely
  to be affiliated.  Part of the observed change can be attributed
  to generational replacement.

\end{itemize}

To estimate the effect of each factor, we simulate a counterfactual
world where the prevalence in the 2010s is the same as in the 1980s.
As a baseline, we use Model 2 to fit, for each respondent, the
probability of having a religious affiliation, and add up the total
probability.

To evaluate the effect of religious upbringing, we use Model 2 and the
data from the 2000s.  We modify \verb"has_relig" for randomly-selected
respondents so that the prevalence of religious upbringing is at the
level seen in the 1980s.  In this counterfactual world, there would be
an additional 6.3 million people with religious affiliation (95\% CI
5.7--7.1), compared to the baseline model.  So changes in religious
upbringing account for 25\% of the observed increase of 25 million.

To quantify the effect of the increase in of college education, we
modify \verb"educ_from_12" to simulate educations patterns from the
1980s.  We find that the increase in college education accounts for
1.3 million of the disaffiliated (95\% CI 0.8--1.9), or 5\% of the
observed increase.

To simulate a world with no Internet use, we set \verb"wwwhr" to 0 for
all respondents.  We find that Internet use accounts for 5.1 million
people with no religious affiliation (95\% CI 1.8--7.2), or 20\% of
the observed increase.

To estimate the effect of generational replacement, we simulate an
earlier cohort by subtracting 20 years from \verb"born_from_1960".  In
this counterfactual world, an additional 13.4 million people would be
affiliated, or 53\% of the observed change.

In summary, the impact of Internet use is comparable to the effect of
religious upbringing, and about 4 times greater than
the effect of college (and post-graduate) education.  In total,
these three effects account for 50\% of the observed increase.

The remainder of the change is accounted for by generational
replacement, although this explanation is unsatisfying because
year of birth cannot, itself, be a causal factor.  So about half
of the observed change remains unexplained.

% upbringing 6.28 (5.67, 7.04)
% upbringing 25.1 (22.7, 28.2)
% college 1.25 (0.722, 1.89)
% college 4.99 (2.89, 7.55)
% internet 5.36 (1.98, 7.66)
% internet 21.4 (7.91, 30.6)
% generation 13.4 (11, 15.8)
% generation 53.4 (44, 63.1)

\section{Discussion}

We have identified three factors that are statistically associated
with religious affiliation: religious upbringing, education and
Internet use.  This association does not prove
causation; if A and B are associated, it is possible that A
causes B, B causes A, or a third factor causes both A and B.

Religious upbringing is by far the strongest explanatory variable, and
it seems likely that, in fact, religious upbringing causes religious
affiliation.  First, it is impossible for religious affiliation as an
adult to cause religious upbringing (although it might color the way
respondents report their upbringing).  Second, it is hard to imagine a
third factor that would cause both.  Finally, it is easy to imagine
the mechanism by which religious instruction could lead to lifelong
affiliation.

For college education, it is less obvious that the relationship is
causal, but we can make several supporting arguments.  Again, it is
not possible that religious affiliation as an adult causes college
graduation in the past.  On the other hand, it is easy to imagine that
religious upbringing might affect both.  We can test that possibility
by running a regression with \verb"college" as the dependent variable
and explanatory variables \verb"had_relig", \verb"top75_income",
\verb"born_from_1960", \verb"sei", \verb"urban", \verb"rural".

We find that the effect of religious upbringing on college graduation
is small, positive, and borderline statistically significant (odds
ratio 1.2, $p=0.065$).  So even if religious upbringing affects
college graduation, it does not explain the {\em negative}
relationship between college education and religious affiliation
or the decrease in religious affiliation over time.
Finally, it is easy to imagine mechanisms by which the experience of
attending college might decrease the chance of religious affiliation.

Similarly, it is easy to imagine at least two ways Internet use could
contribute to disaffiliation.  For people living in homogeneous
communities, the Internet provides opportunities to find information
about people of other religions (and none), and to interact with them
personally.  Also, for people with religious doubt, the Internet
provides access to people in similar circumstances all over the world.
Conversely, it is harder (but not impossible) to imagine plausible
reasons why disaffiliation might cause increased Internet use.

Again, it is possible that religious upbringing might have a causal
connection with Internet use, yielding a spurious relationship
between Internet use and affiliation.
We can test that possibility by running a
regression with \verb"www2" as the dependent variable and explanatory
variables \verb"had_relig", \verb"top75_income",
\verb"born_from_1960", \verb"sei", \verb"urban", \verb"rural".  We
find that the effect of religious upbringing on Internet use is small,
positive, and {\em not} statistically significant (odds ratio 1.1,
$p=0.37$).  It is unlikely that religious upbringing has a negative
effect on Internet use large enough to explain the
relationship between Internet use and affiliation.

Although a third unidentified factor could cause both disaffiliation
and Internet use, we have controlled for most of the obvious
candidates, including income, education, socioeconomic status, and
rural/urban environments.  Also, in order to explain changes over
time, this third factor would have to be new and rising in
prevalence, like the Internet, during the 1990s and 2000s (see
Figure~\ref{fig.internet}).  It is hard to imagine what that factor
might be.

\section{Conclusions}

Someone who has taken an introductory statistics class might insist
that correlation does not imply causation, and that is a useful
reminder.  Nevertheless, correlation does provide evidence in
favor of causation, especially when we can eliminate alternative
explanations or have reason to believe that they are less likely.

So until there is another explanation for the associations reported
here, it is reasonable to conclude, at least tentatively:

\begin{itemize}

\item Religious upbringing increases the chance of religious
affiliation as an adult.  Decreases in religious upbringing
between the 1980s and 2000s account for about 25\% of the
observed decrease in affiliation.

\item College education decreases the chance of religious
affiliation.  Increases in college graduation between the
1980s and 2000s account for about 5\% of the
observed decrease in affiliation.

\item Internet use decreases the chance of religious
affiliation.  Increases in Internet use since 1990, from 0 to nearly
80\% of the general population, account for about 20\% of the
observed decrease in affiliation.

\end{itemize}

The remaining 50\% of the decrease in religious affiliation is
accounted for by generational replacement, but this ``explanation''
only raises the question of why, even after accounting
for education and Internet use, people born later are more
likely to disaffiliate.

\section{Limitations}

The GSS data imposes some limitations on our analysis.

\begin{itemize}

\item In the cumulative GSS dataset, annual income is reported in 11
  bins between \$1000 and \$25000, and one bin for higher incomes.
  Since 75\% of respondents fall into the highest bin, we cannot
  measure the effect of higher incomes, if any.

  However, we used GSS individual year datasets to obtain income
  data with a wider range, and checked whether income in the
  other quartiles had an effect on disaffiliation.  We tested variables
  set at the 50th and 75th percentiles of income, but found that
  neither contributed an effect that was statistically
  significant, and adding these variables had almost no effect
  on the coefficients of the other variables.

  Other studies have found the same relationship: people in low
  income brackets are less likely to have a religious affiliation,
  but there are no statistical differences among the other
  brackets\cite{pew2008us}.

  We conclude that the income variable we used is sufficient to
  control for the effect of income on affiliation.

\item By design, the GSS does not ask all questions to all
  respondents.  Out of 14 948 respondents only 8951 were asked and
  answered all questions included in our Models 1 and 2.  This
  selection process introduces some small biases.  Respondents
  with complete records were slightly less likely to have a
  religious affiliation (84.2\% compared to 84.9\%), more
  likely to be in the top three income quartiles (75.3\% compared
  to 73.0\%), more likely to have graduated from college
  (31\% compared to 27\%), and about a year older.

  Although the selected sample differs in some ways from the general
  population, it is unlikely that this selection bias has a
  substantial effect on the relationships reported in this paper.

% all respondents
%has_relig 0.849066045486
%had_relig 0.924719445087
%top75_income 0.730185055664
%mean born_from_1960 -0.939318141136
%college 0.271886111959

% complete records
%has_relig 0.842657580086
%had_relig 0.921307639474
%top75_income 0.752809964367
%mean born_from_1960 0.0828757203717
%college 0.310006111401

\item In 2010 the GSS used a new protocol to screen questions about
  Internet use.  As a result, it is not possible to analyze changes in
  Internet use after 2006.  The same protocol is being used in 2012,
  so the next data point for time series analysis will be available in
  2015 at the earliest.

\end{itemize}

\section{Methodological notes}

Confidence intervals and p-values in this paper are estimated by
resampling, which relies on minimal assumptions.  
We draw random samples from the observed sample, using respondent
weights so that each generated sample is representative of the general
population.  We compute regressions for 1001 generated samples and
compute the median, 95\% confidence interval, and p-value for each
parameter.

To compare models, we use the self-information of partitioning (SIP)
to compute the total surprisal of each model.  For example,
if the model predicts that the probability of the dependent
variable for the $i$th respondent is $p_i$ and we learn that
the actual value is 1, we have learned $-log_2(p_i)$ bits of information.
If we learn that the actual value is 0, that's
$-log_2(1-p_i)$ bits.

So if we compute fitted values of $p_i$ and then learn the actual data, the
total number of bits we learn is:

\[ -\sum x_i \log_2(p_i) + (1-x_i) \log_2(1-p_i)\]

With a perfect model $p_i = x_i$ and the total information of the data
is 0.  As the quality of the model decreases, the information
remaining in the data increases.  So the better model is the one
that leaves less information in the data.

To adjust for the number of parameters, we compare each model to a
baseline that contains the same number of parameters, but where the
explanatory variables have been replaced with random values.

Details of these methods are in the comments and code available from
\url{https://github.com/AllenDowney/internet-religion}.

\bibliographystyle{plain}
\bibliography{paper}

\end{document}